\documentclass[pra,aps,superscriptaddress,twocolumn,notitlepage,showpacs]{revtex4-2}

\usepackage{amssymb,mathrsfs,bm,color,times}
\usepackage{graphicx,color}
\usepackage{CJK}
\usepackage{bm}
\usepackage{amsmath}
\usepackage{graphicx} 
\usepackage{epstopdf}

\newcommand{\be}{\begin{equation}}
\newcommand{\ee}{\end{equation}}
\newcommand{\bea}{\begin{eqnarray}}
\newcommand{\eea}{\end{eqnarray}}
\newcommand{\bsube}{\begin{subequations}}
\newcommand{\esube}{\end{subequations}}

\newcommand{\beq}{\begin{equation}}
\newcommand{\eeq}{\end{equation}}
\newcommand{\beqn}{\begin{eqnarray}}
\newcommand{\eeqn}{\end{eqnarray}}

\newcommand{\bsub}{\begin{subequations}}
\newcommand{\esub}{\end{subequations}}

\def\ft#1#2{{\textstyle{\frac{\scriptstyle #1}{\scriptstyle #2} } }}
\def\fft#1#2{{\frac{#1}{#2}}}

\begin{document}

\title{Two-dimensional UV femtosecond stimulated Raman spectroscopy for molecular polaritons: dark states and beyond}

\author{Jianhua Ren}
\affiliation{Department of Physics,
  City University of HongKong, Kowloon, Hongkong SAR}

\author{Zhedong Zhang}
\email{zzhan26@cityu.edu.hk}
\affiliation{Department of Physics,
  City University of HongKong, Kowloon, Hongkong SAR}
\affiliation{
  City University of HongKong, Shenzhen Research Institute, Shenzhen 518057, Guangdong, China}

\date{\today}

\begin{abstract}
We have developed a femtosecond ultra-voilet (UV) stimulated Raman spectroscopy (UV-FSRS) for $N$ molecules in optical cavities. The scheme enables a real-time monitoring of collective dynamics of molecular polaritons and their coupling to vibrations, along with a crosstalk between polariton and dark states.  Through multidimensional projections of the UV-FSRS signal, we identify clear signature of the dark states, e.g., pathways and timescales that used to be invisible in resonant technique.  A microscopic theory is developed for the UV-FSRS, so as to reveal the polaritonic population and coherence dynamics that interplay with each other. The resulting signal makes the dark states visible, thereby providing a new technique for probing dark state dynamics and their correlation with polariton modes.

\end{abstract}


\maketitle

\section{Introduction}
The strong interactions between photons and molecules in micro-cavities lead to hybrid phases of matter, forming a superposition of molecular states and photons \cite{polariton}.  These  excitations, known as molecular polaritons, may present unusual properties incredibly distinct from normal molecules, for instance, the controllable many-body couplings and the cooperative emission of light.  The complexity of molecules, due to the various degrees of freedom, has resulted in rich interactions between the excitations subject to multiple scales. A variety of intriguing phenomenon was therefore reported in recent studies, including polariton lasing \cite{76-18} and condensation \cite{76-19,20-3}, cavity-altered reactivity of chemistry  \cite{20-4,20-5,20-6,20-7,42-30,42-33,42-32} and topological effects  \cite{42-31}. All these highlight the importance of the polariton dynamics, which however remains elusive. 

So far, the strong coupling of molecules to cavities has led to extensive studies along with intense debates arising from the dark states. In spectroscopy, such kind of states is hard to be visualized, e.g., absorption and fluorescence. Nevertheless, the nonradiative processes—the channels causing symmetry breaking but not existing in atomic ensembles—may lead to the energy/information leakage from polariton states. Much theoretical and experimental efforts have been devoted to the relaxation of polariton modes, whereby the dark states serve as exciton reservoirs for the optical systems. Notably, the crosstalk between molecular polaritons and pure molecular modes was observed in recent experiments  \cite{JPCA1235918,NC}. A selective dynamics of dark-state polaritons coupled to bright polaritons was therefore demonstrated. Owing to the coherent and invisible nature, the dark-state polaritons show a dephasing presumably crucial for the energy transfer process  \cite{MarkusJPCA}. Moreover, the high mode density makes the dark states a good strategy for controlling the chemical reactivity and achieving the phase transition towards the polariton condensation 
 \cite{PRL128096001,PRB106L220306}. Elaborate experiments demonstrated unusual dynamics of molecular polaritons beyond the Tavis-Cummings model, when considering the condensed-phase molecules  \cite{cd1,cd2}. In the presence of solvent-induced disorder, extensive studies showed the spectral lines as a signature of the pure molecular states weakly coupled to cavity photons  \cite{ZhangJPCL2019, XiongPNAS2018}. This indicates the localization nature. All these call for a comprehensive understanding of the polariton dynamics in a conjunction with the dark states in molecules.

In this article, we propose a novel off-resonant spectroscopic probe for the molecular polaritons, based on the stimulated Raman scattering. A two-dimensional ultraviolet femtosecond stimulated Raman spectra (2DUV-FSRS) is developed for molecules strong interacting with microcavities. Using a combination of visible pump and UV probe pulses, the dark-state polaritons show a prominent Raman response. Our results demonstrate a multi-dimensional projection of the coherent Raman signal for a real-time monitoring of the dark-state dynamics. A microscopic polariton model is further developed for the 2DUV-FSRS, elaborating the multi-timescale nature for the population dynamics of polaritons in a crosstalk with the dark states.

The rest of the paper is organized as follows. In Section \ref{sec: model}, we discuss the dynamics of the model used to describe the interactions between photons and molecules. Section \ref{sec: preliminary} provides a brief review on the Raman spectroscopy. In Section \ref{sec: onedimen}, we derive and numerically present the one-dimensional Raman signal in real time for the model under consideration. Besides, we propose a resolution procedure that allows us to quantitatively extract the dynamics of both polaritons and dark states. We then analyze the two-dimensional Raman signal, taking into account the time delay between pulse $\varepsilon_1$ in Section \ref{sec: result}. Furthermore, we have discussion on the charge transfer
state in Section \ref{sec: ct}. Finally, Section \ref{sec:conclu} presents our conclusions and remarks.

\section{Model for molecular excitons}
\label{sec: model}

\subsection{Polariton and Dark state}

\begin{figure}[h]
\includegraphics[width=0.55\textwidth]{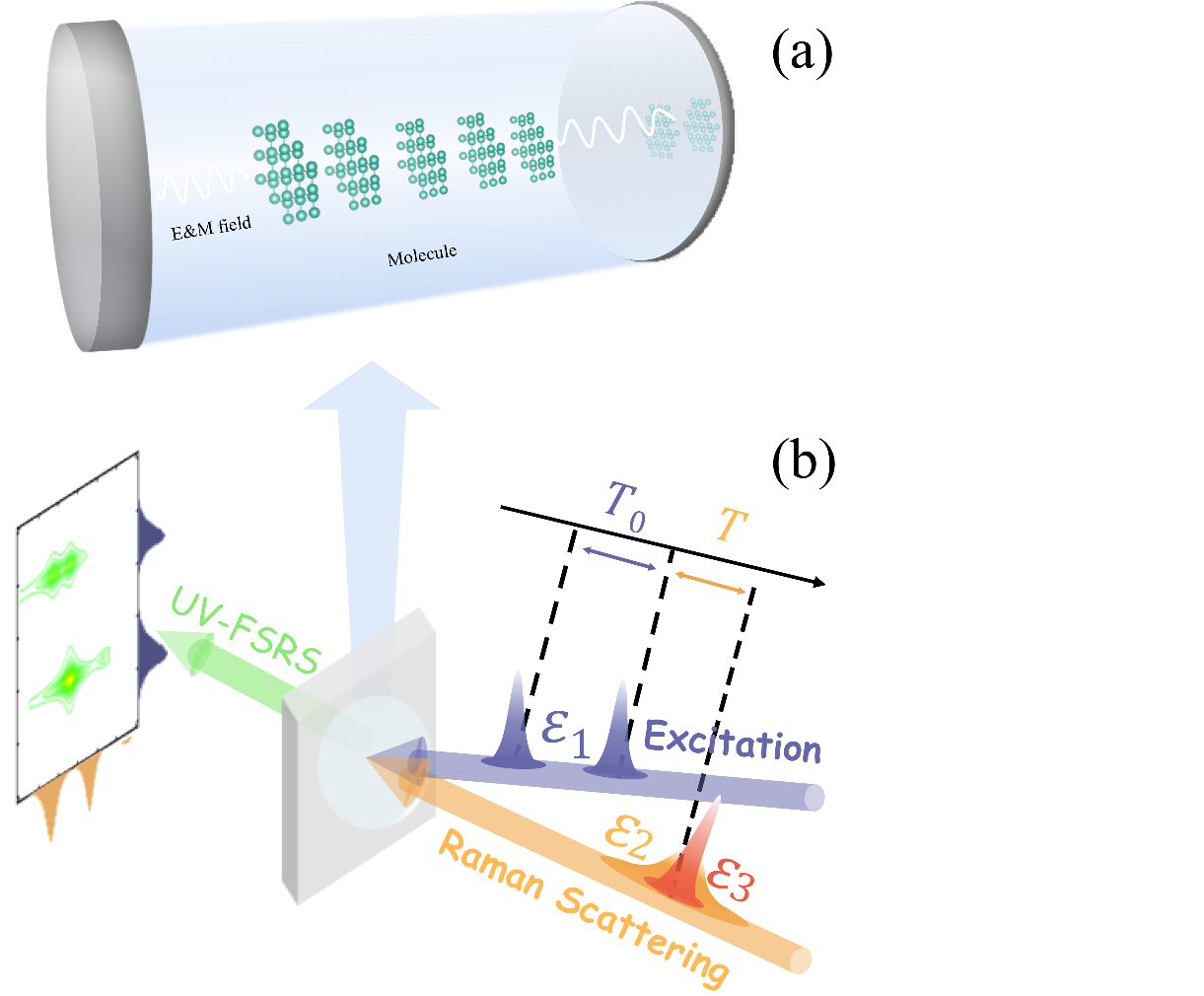}
\caption{(a) Microscopic structure of polariton. (b) Schematic illustration of UV-FSRS setup. }
\label{mp}
\end{figure}

We consider a generic model consisting of N molecules in a single-mode optical cavity, as depicted in FIG. \ref{mp}(a). Each molecule has one exciton mode, describing the electronic excitations. The excitons essentially interact with the vibrations that may feature a dense distribution in complex molecules. This system can be described by a model Hamiltonian $H=H_0 + H_{\rm vib} + V_{\rm int}$, where $H_0$ is the exciton-photon component, e.g.,

\begin{equation}
H_0=\sum_{j=1}^{N}\left[\omega b_{j}^{\dagger} b_{j}+g\left(b_{j}^{\dagger} a+b_{j} a^{\dagger}\right)+\frac{\mathcal{U}}{2} b_{j}^{\dagger} b_{j}^{\dagger} b_{j} b_{j}\right]+v a^{\dagger} a
\label{1e}
\end{equation}

\begin{figure}[h]
\includegraphics[width=0.25\textwidth]{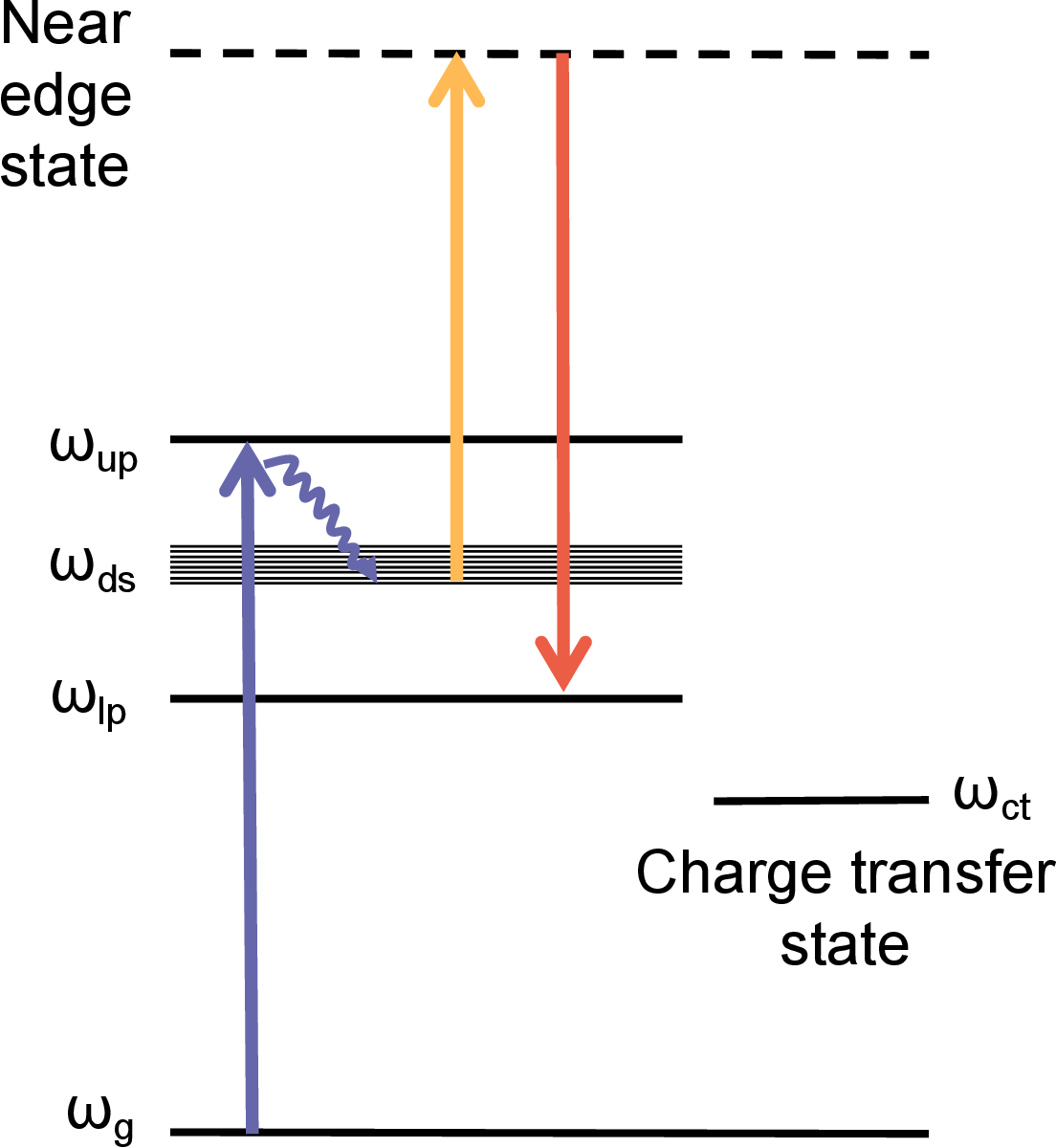}
\caption{An illustration of the energy levels in the cavity, including the upper-polariton $\omega_{\rm up}$, dark state $\omega_{\rm ds}$, lower-polariton $\omega_{\rm lp}$, the ground state $\omega_{\rm d}$, as well as near edge state and charge transfer state $\omega_{\rm ct}$.}
\label{energylevel}
\end{figure}

Here, $b_j$ and $b_j^\dagger$ are the respective annihilation and creation operators for excitons in the $j$th molecule, satisfying $[b_i,b_j^\dagger]=\delta_{ij}$.  $a$ and $a^\dagger$ denote the annihilation and creation operators for photon modes with $[a,a^\dagger]=1$. $\omega$ and $v$ are the energy of the single mode for the molecule and the cavity, respectively.  The photon-molecule coupling is denoted by $g={-\sqrt{{2 \pi \hbar v}/{V }}{p} }$, $V$ is cavity volume, $p$ is dipole moment of molecule.   And the parameter $\mathcal{U}=\sum_{a,b}\langle j_a,j_b|({\vec{d}_{a} \cdot \vec{d}_{b}-3\left(\hat{n} \cdot \vec{d}_{a}\right)\left(\hat{n} \cdot \vec{d}_{b}\right)})/{4 \pi \epsilon_0 R^3}|j_a,j_b\rangle$ measures the interaction strength between excitons. where $\vec{{d}}_i=e \vec{{r}}_i$ are the exciton dipole moments in one molecule and $|j_a\rangle$ reprents the exciton state a in jth molecule. $R$ is the separation between the dipoles and $\hat{n}=\vec{R} / R$ the unitary vector in the direction from one dipole to another \cite{u}. One observes the total number of excitations, i.e., $M=\sum_{j=1}^{N} b_{j}^{\dagger} b_{j}+a^{\dagger} a$ which is conserved due to $[M,H_0]=0$. The Hamiltonian in Eq.(\ref{1e}) is thus of a block-diagonal form. This leads to a superposition subject to a certain M, forming new molecule-cavity states. In the absence of $\mathcal{U}$, two modes called upper polariton (UP) and lower polariton (LP) are found, with the other N-1 dark states (DS) or the so-called dark-state polaritons (DSPs). A schematic diagram for the energy structure is demonstrated in FIG. \ref{energylevel}, where we denote the energy of UP, LP and DSPs as $\omega_{\rm up}$, $\omega_{\rm lp}$ and $\omega_{\rm ds}$ respectively, in addition to the trivial ground state $\omega_{\rm g}$; we also include near edge state and charge transfer state, which exists to trigger different Raman processes.



The coupling of excitons to vibrational modes is of the form  \cite{75}
\begin{equation}
    V_{\rm int}= \sum_i^N \sum_k {\omega^i_{{\rm{vib}},{k}}}q_{i,k} d_{i,k}b_i^\dagger b_i
\end{equation}
and $H_{\rm vib}=\sum_{k} \sum_{i}^{N}\frac{1}{2} \omega_{{\rm vib},k}^i C_{i k}^{\dagger} C_{i k}$ where the vibrational modes are assumed to have a dense distribution that can be described by a smooth spectral density. $q_{i,k}=\sqrt{{\hbar}/{2m{\omega_{{\rm vib},k}^i}}}(C^\dagger_{i,k}+C_{i,k})$ denotes the coordinates of the molecular stretching. Defining the polariton operators
\begin{equation}
    \begin{split}
        \eta_n = \sum_j^N U_{n,j} b_j + U_{n,N+1} a
    \end{split}
\end{equation}
through diagonalizing the $H_0$ apart from the $\mathcal{U}$ term in Eq.(\ref{1e}), we have $[\eta_n,\eta_m^{\dagger}] = \delta_{nm}$ and recast the Hamiltonian into
\begin{subequations}
   \begin{align}
    & H_0 = \sum_{i=1}^{N+1} \omega_{ i} \eta_i^{\dagger} \eta_i+\fft{\mathcal{U}}{2} 
    \sum_{k, l, m, n}^{N+1} K_{klmn} \eta_k^{\dagger} \eta_l^{\dagger} \eta_m \eta_n \\[0.15cm]
    & V_{\rm int}= \sum_{m>n,i,k}^{N+1}
    \mathcal{V}_{i,k}
    \left(U_{im}^\dagger U_{ni}\eta^\dagger_m\eta_nC_{i,k} + \text{h.c.} \right)
  \end{align}
\end{subequations}
where $\omega_{N+1/N}=\omega_{\rm up/lp}=\frac{1}{2}\left(v \pm \sqrt{4 g^2+(v-\omega)^2}+\omega\right) $, $\omega_{1,2..N-1}=\omega_{ds}=\omega$, $K_{k l m n}= \sum_j^N U^{\dagger}_{j k} U^{\dagger}_{jl} U_{mj} U_{ n j}$, $\mathcal{V}_{i,k}=\sqrt{{1}/{2}}\omega_{{{\rm{vib}},k }}^i d_{i,k}$.
Note that we have not included terms such as ${\eta_m^\dagger\eta_nC_{i,k}^\dagger}$because they are negligible according to the rotating wave approximation.

Since the number of intramolecular vibrational degrees of freedom is significantly large and their interaction with molecules is notably weak, we can treat the vibrations as a bath system which affects the molecule-photon system with negligible backreactions. Therefore, we consider the molecule-photon cavity system as an open system by tracing out the vibrations and obtain the polariton Redfield equation $\dot{\rho}=-i[H_0,\rho]+\hat{\mathcal{W}}\rho$  \cite{chem2009}
\begin{subequations}
   \begin{align}
    & \hat{\mathcal{W}}{\rho}=\sum_{m>n} \frac{\gamma_{mn}}{2}[(\eta_m^\dagger\eta_n\rho \eta_n^\dagger\eta_m-\eta_n^\dagger\eta_m\eta_m^\dagger\eta_n\rho)\bar{n}_{w_{m n}} \nonumber\\
& + (\eta_n^\dagger\eta_m\rho\eta_m^\dagger\eta_n-\eta_m^\dagger\eta_n \eta_n^\dagger\eta_m \rho)(\bar{n}_{\omega _{m n}}+1)]
+ \text{h.c.}
 \nonumber\\
& \label{master}
\end{align}
\end{subequations}
with $\gamma_{mn}=\sum_i^NJ_i(\omega_{mn})U_{mi}^* U_{mi} U_{ni} U_{ni}^*$ and the spectral density  \cite{gamma}
\begin{equation}
J_i(\omega_{mn})
=2\lambda_0\fft{\omega \gamma_0}{\omega^2+\gamma_0^2}.
\end{equation}

The solution to the polariton Redfield equation is given by
\begin{equation}
    \rho_{{e_{4}} {e_{3}}}(t)=\sum_{{e_{2}}{ e_{1}}} \mathrm{G}_{{e_{4}}{ e_{3}},{ e_{2}}{ e_{1}}}(t) \rho_{{e_{2}}{ e_{1}}}(0) \,
\label{mastersolution}
\end{equation}
where $G(t)$ is the Green's propagator for Eq.\eqref{master}, in the absence of external fields  \cite{chem2009}. The polariton dynamics governed by Eq.(\ref{mastersolution}) will be imprinted into the Raman response, when interacting with laser pulses. The nonlinear optical signals are thus capable of reading out the polariton resonance and dynamics.

\section{Preliminary on Raman spectroscopy}
\label{sec: preliminary}

In this work, our strategy involves applying stimulated Raman techniques to study femtosecond ultraviolet stimulated Raman spectroscopy (UV-FSRS) for molecular polaritons and optically-dark states  \cite{2,3,5,6,8}. 
We choose the UV light, as the UV-light-induced Raman transition would be background-free, due to the off-resonant nature. Moreover, the UV laser pulses can make it feasible to induce the electronic Raman polarizability via the near-edge states of molecules.

For the 2DUV-FSRS, the excitations of system are created by a resonant pump pulses; then a pair of overlapped broad- and narrow-band pulses scatters off the system, so as to produce the stimulated Raman transition. The interaction thus reads
\begin{align}
H_{\rm int} = \alpha \varepsilon_2^\dagger(t)\varepsilon_3(t) - \mu\varepsilon_1^\dagger(t)+ {\rm H.C}
\end{align}
where $\mu$ represents electric dipole operator and $\alpha$ denotes the Raman polarizability operator, i.e., $\alpha = \sum_{e,e'}\alpha_{ee'}|e\rangle\langle e'|$  \cite{raman pro}
\begin{equation}
    \alpha_{{e}{e'}}=\sum_{i}^{N} \frac{P_{i}^2|U_{i {e}}^\dagger U _{{e'}i}|}{\hbar}\left(\frac{1}{\omega_{i}-\omega_{{e'}}}+\frac{1}{\omega_{i}-\omega_{{e}}}\right).
\label{35}
\end{equation}
The energy of the near-edge states $|r_i\rangle$ is high enough to ensure that it does not admit any interactions with the cavity  \cite{near}, and $P_i$ refers to the transition dipole between different energy levels in the $i$th molecule, namely $P_i=\langle e_i|\mu|r_i\rangle$.
 Since $\left|r_{j}\right\rangle$ is highly excited and thus is decoupled from the cavity, a random phase is essentially attached to the Raman transition amplitude, i.e., $U_{e i}=|U_{ei}|e^{i \phi_{ei}}$. Therefore, we have to take the ensemble average over the phase $\langle |U^{\dagger}_{ie}U_{e' i}|e^{i(\phi_{e'i}-\phi_{ei})} \rangle_{\rm ave}=  |U^{\dagger}_{ie}U_{e' i}|$, ensuring that Raman polarizability between two excited states is well-defined.

The transmission of the short pulse (Raman process) is measured, so that the optical signal is $S= \langle \varepsilon_3^*(t) \varepsilon_3(t)\rangle$. Using the Heisenberg's equation of motion for fields one has  \cite{fsrs}:
$
S=\int d t\langle(-i)[\varepsilon_3^{\dagger}(w) \varepsilon_3(w), H_{int} ]\rangle, 
$
where the most significant terms are remained, i.e., the resonant pump and Raman probe that are time ordered. As the Raman pulses excite the molecules, polariton modes then begin to interact with both dark states and the vibrational bath. This interaction causes relaxation and repopulation, and the relevant dynamical information is encoded in the density matrix that can be described by the Born-Redfield equation  \cite{chem2009}. As a result, the multidimensional projections of the UV-FSRS signal reflect the real-time information regarding the dark states and their correlations with polariton modes. More specifically, the Raman signal can be evaluated by:
\begin{equation}
S=\frac{1}{\pi} \Im \int d t e^{i \omega(t-T)} \varepsilon_3^*(w) \varepsilon_2(t-T) \cdot \operatorname{Tr}[\hat{\alpha} \rho]\label{eq: signal}
\end{equation}
In this work, we consider density matrix $\rho$ at three order.

\begin{figure}[h]
\includegraphics[width=0.35\textwidth]{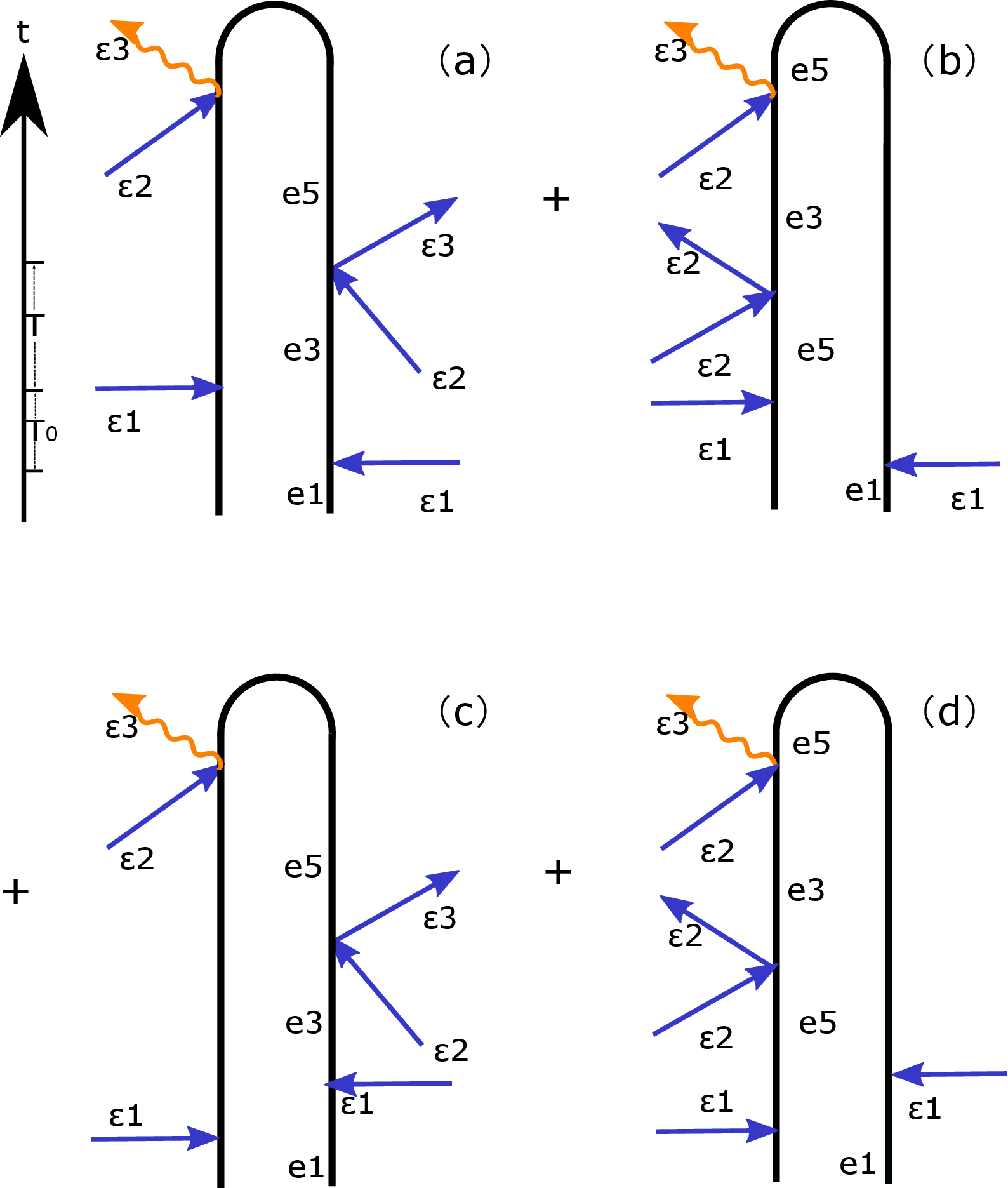}
\caption{loop diagram of raman spectroscopy}
\label{loop}
\end{figure}

\section{One-dimensional UV femtosecond stimulated Raman spectroscopy}

\label{sec: onedimen}

To calculate the Raman signal from eq.\eqref{eq: signal}, we must perform the integral over the time domain. This is hard in general, but can be achievable by assuming the Lorentzian pulse shape,
\begin{equation}
     \varepsilon_i(t-T_i)=\theta(t-T_i) e^{-\frac{t-T_i}{\sigma_i}}e^{-i\omega_i t}
\end{equation}
where $T_i$ and $\omega_i$ denote the central time and frequency of the pulse, respectively. The resonant pump $\varepsilon_1$ excites the system, and the pulses $\varepsilon_2$ and $\varepsilon_3$ act with a time delay $T=T_2-T_1$ relative to $\varepsilon_1$, inducing the Raman transition.  The pulse fields inducing Raman transition have identical arrival time, i.e.,$T_2=T_3$. Due to the broadband nature of the resonant pump, a wide spectrum of the excited states is covered. The Raman signal given by Eq.(\ref{eq: signal}) thus includes populations $\rho_{ee}$ and coherence $\rho_{ee'}$ ($e\neq e'$) components of molecular polaritons. 

When the Raman fields and the pump field are well separated in time, namely, $T\gg \sigma_1, \sigma_2, \sigma_3$, the FSRS spectroscopic experiment can be viewed as a three-step process: preparation, propagation and Raman detection. The two resonant pump fields create an initial doorway state of the polaritons, which propagates, and is finally probed at a time delay $T$ with a window operation. For a precise definition of the three-step process, we expand Eq.(\ref{eq: signal}) against the Raman coupling, having
\begin{equation}
   \begin{split}
           S_{\text{1D}} & (\omega - \omega_2,T) \propto N \int_{-\infty}^{\infty} dt \int_{-\infty}^{t} d\tau e^{i(\omega-\omega_2)t} \varepsilon_3^*(\omega) \\[0.2cm]
    & \times \varepsilon_2(t-T) \varepsilon_2^*(\tau-T)\varepsilon_3(\tau-T) \text{Tr} \Big\{ \alpha \big[\alpha(\tau), \rho_0\big] \Big\} \\[0.2cm]
    & \approx N \int_{-\infty}^{\infty} dt \int_{-\infty}^{T} d\tau e^{i(\omega-\omega_2)t} \varepsilon_3^*(\omega)  \varepsilon_2(t) \varepsilon_2^*(\tau) \\[0.2cm]
    & \qquad\qquad\qquad \times \varepsilon_3(\tau) \text{Tr} \Big\{ \alpha(t) \big[\alpha(\tau), \rho(T)\big] \Big\}.
   \end{split}
\label{S1Dp}
\end{equation}
with a proper approximation in last step, given the delay $T$ longer than the pulse duration. Such an approximation works for most of the ultrafast molecular spectroscopic experiments. This allows the definition of the {\it Raman window operators}
\begin{subequations}
    \begin{align}
        & W(\omega-\omega_2) = \int_{-\infty}^{\infty} dt e^{i(\omega-\omega_2)t} \varepsilon_3^*(\omega) \varepsilon_2(t) \alpha(t), \label{RW} \\[0.2cm]
        & V(\sigma) = \int_{-\infty}^T d\tau \varepsilon_2^*(\tau) \varepsilon_3(\tau) \alpha(\tau). \label{RV}
    \end{align}
\end{subequations}

Using Eqs.(\ref{S1Dp}), (\ref{RW}) and (\ref{RV}), the UV-FSRS signal for the Raman shift $\omega-\omega_2$ reads
\begin{equation}
    S_{\text{1D}} (\omega - \omega_2,T) \propto \Re \text{Tr} \Big\{W(\omega - \omega_2) \big[V(\sigma),\rho(T) \big] \Big\}.
\label{1dp}
\end{equation}

To prepare the polariton state $\rho(T)$ the {\it doorway operators} are useful to describe the process by the resonant pump pulses. A {\it doorway-Raman-window} representation can be therefore developed, which would be powerful for an unified understanding of the multidimensional Raman spectroscopy. We will elaborate this in Sec.\ref{sec: result} for the two-dimensional UV-FSRS.

\begin{figure}[h]
\includegraphics[width=0.45\textwidth]{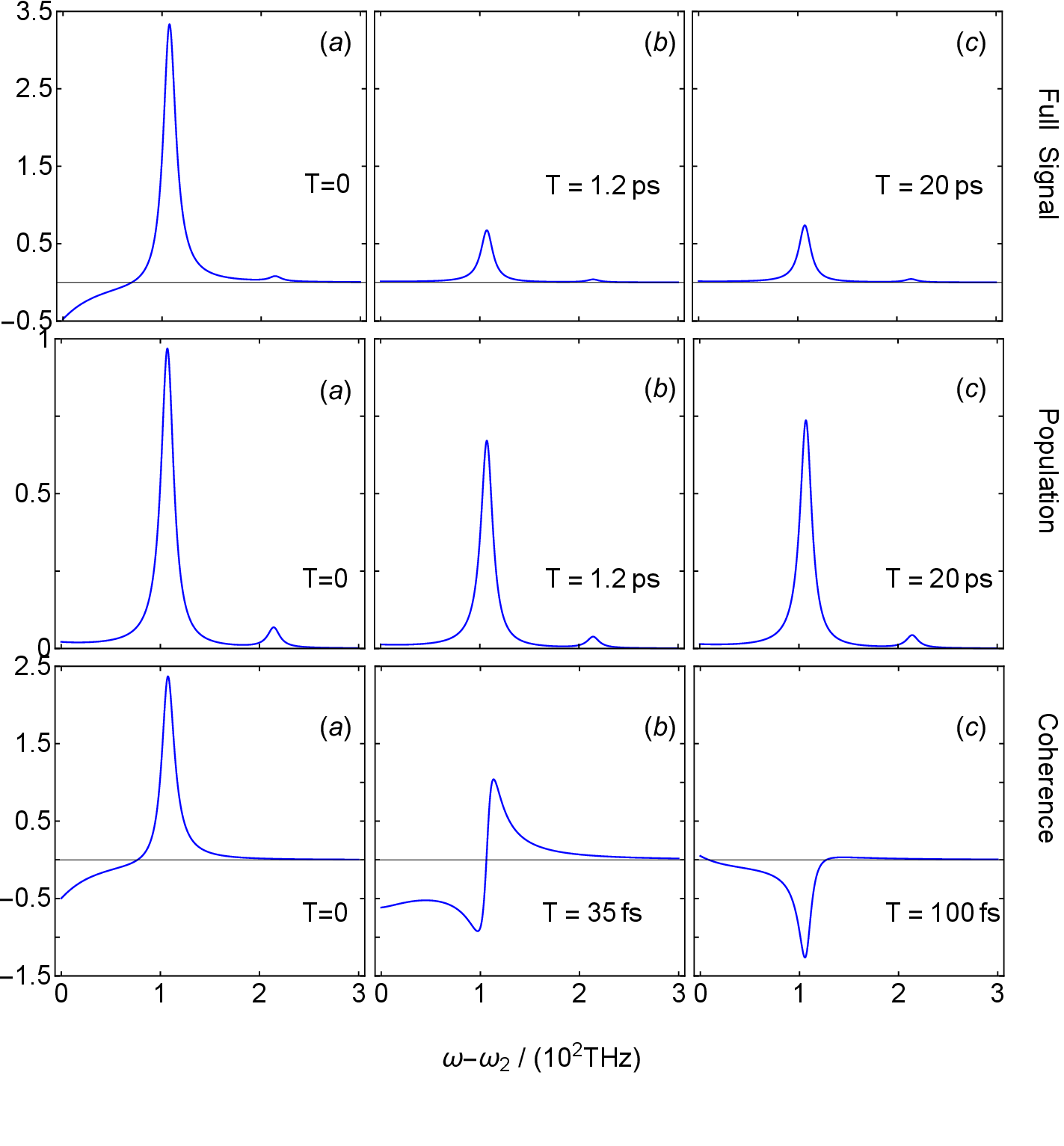}
\caption{The stimulated Raman signals of $N=10$ molecules, without detuning, are shown. In this paper, we consider cyanine dye J-aggregates \cite{jagg} with $\omega=1.84 {\rm ev}$ and $\mathcal{U}=0.02\rm ev$, with $\hbar=1$ throughout our discussions. We have taken $\delta=243$THz as the center frequency difference, $\sigma_2=1$ps, and $\sigma_3=35$fs for the broad pulse $\varepsilon_2$ and the narrow pulse $\varepsilon_3$, respectively. To better simulate the real situation, we introduce the phenomenological parameter $\gamma=100$THz to modify $\xi$s by $\xi_{e_5e_3}=\omega_{e_5e_3}- i (\gamma_{e_5e_3}+\gamma)$, which can account for the possible dephasing rate caused by experimental uncertainty. Subsequent signals use the same parameters. Horizontally, (a), (b), and (c) monitor the real-time evolution of the stimulated Raman signals with different time delays $T$; vertically, the total signals, populations, and coherence parts are shown, respectively. The relative strength of peaks observed in the real-time delay can be used to resolve the populations of molecular states.}
\label{1d0}
\end{figure}

To see the Raman signal closely and study the dynamics of polaritons and dark states numerically, we set a strong coupling $g \sqrt{N}=0.05\sqrt{2}$ without losing generality. We have simulated the polariton FSRS signals (given by Eq. \eqref{1dp} ) in FIG. \ref{1d0} and FIG. \ref{1dtot+} with different detunings.

Before moving on to the detailed analysis of the signals, let us briefly comment on the implications derived from the signal in one dimension without referring to many numerical details. According to first-order perturbation theory, we know that the pulse $\varepsilon_1$ initially excites the polaritons to the superposition state $\mathcal{N}^{-1}(\mu_{\rm up,g}|\rm UP\rangle+\mu_{\rm lp,g}|\rm LP\rangle)$, where $\mathcal{N}=\sqrt{\mu^2_{\rm up,g}+\mu^2_{\rm lp,g}}$ is the normalization factor.
Generally speaking, the intensity of the signal curve indicates the probability of a certain process occurring, which boils down to the population of the relevant states. For example, a peak located at $\omega_{e_3e_5}\geq 0$ reflects that the linear combination of density matrices $\rho_{e_3,e_3}+\rho_{e_5,e_5}$ is dominating, rather than the individual population $\rho_{e_3,e_3}$ or $\rho_{e_5,e_5}$. This is attributed to the two components: dissipative (FIG. \ref{loop}(a,c)) and parametric (FIG. \ref{loop}(b,d)) processes that give the same Raman resonance. This argument can guide us to qualitatively decode more populations from the Raman signal.

In the following part, one-dimensional UV-FSRS will be discussed in detail for cases with or without detuning between molecules and the cavity, under specific parameters.

\subsection{Zero cavity-molecule detuning}
In this subsection, we begin with considering the case of zero detuning between the molecular excitons and the cavity, i.e., $\omega-v=0$. In FIG. \ref{1d0}, we display the total signal, population, and coherence in the top, middle, and bottom lines, respectively.

\begin{figure}[h]
\includegraphics[width=0.35\textwidth]{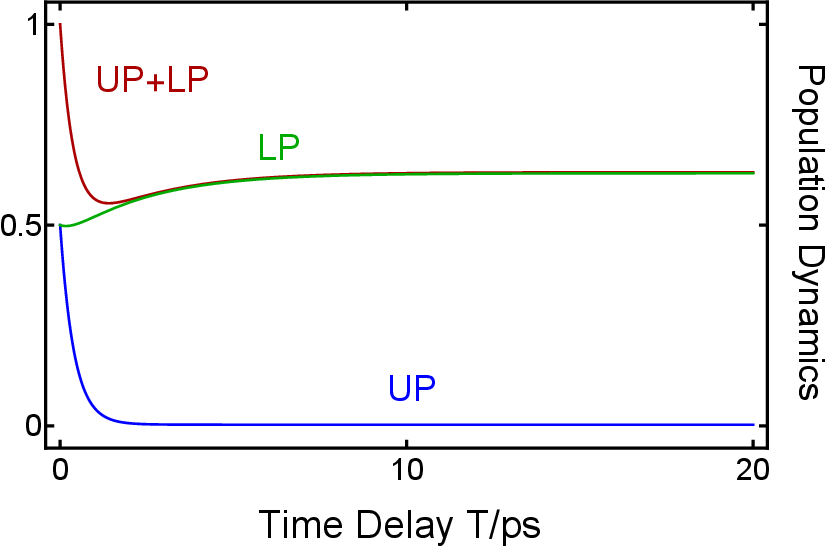}
\caption{The dynamics of populations, solved using the master equation without detuning between the molecules and the cavity. In this case, the molecules are initially pumped to the superposition of the upper polariton and the lower polariton, with the transition dipole being $\mu_{\rm g,up}=\mu_{\rm g,lp}$.}
\label{1dmas0}
\end{figure}

Let's begin with analyzing the population, as shown in the middle line of FIG. \ref{1d0}. Owing to the vanishing detuning $\omega-v=0$, we have $\omega_{\rm up,ds}=\omega_{\rm ds,lp}$. As a consequence, we can observe two peaks: (1) the transitions UP $\leftrightarrow$ Dark and Dark$\leftrightarrow$ LP that merge together; (2) the transition UP $\leftrightarrow$ LP. By varying the time delay T, the populations at the dark states can be resolved from the two peaks in FIG. \ref{1d0}(a,b,c). However, due to the transitions UP $\leftrightarrow$ Dark and Dark $\leftrightarrow$ LP that merge in one peak at 107THz, FIG. \ref{1d0} enables a real-time monitoring of a coupled dynamics for the polariton states, i.e., $\rho_{\rm{up,up}} + \rho_{\rm{lp,lp}}$. This is further confirmed by the simulations of polariton Redfield equation in Eq.(\ref{master}) that gives the population dynamics of the polariton states depicted in FIG. \ref{1dmas0}.

Moreover, FIG. \ref{1d0} shows that the signal resolves the coherence nature of the polaritons, within a short timescale. In particular, the peak at $\omega_{\rm{up,ds}} (\omega_{\rm{ds,lp}})$ presents an oscillation during $\sim$ 1ps, on top of the population components $\rho_{\rm{up,up}} + \rho_{\rm{ds,ds}}$. This reveals the coherence between the UP and the LP.

\subsection{With detuning $\omega-v=1.25g$}

We consider a detuning of $\omega-v=1.25g$ between molecular excitons and cavity photons. As seen from the energy-level structure of polaritons, the two transitions UP $\leftrightarrow$ Dark and Dark $\leftrightarrow$ LP lead to a splitting of the spectral lines.

\begin{figure}[h]
\includegraphics[width=0.4\textwidth]{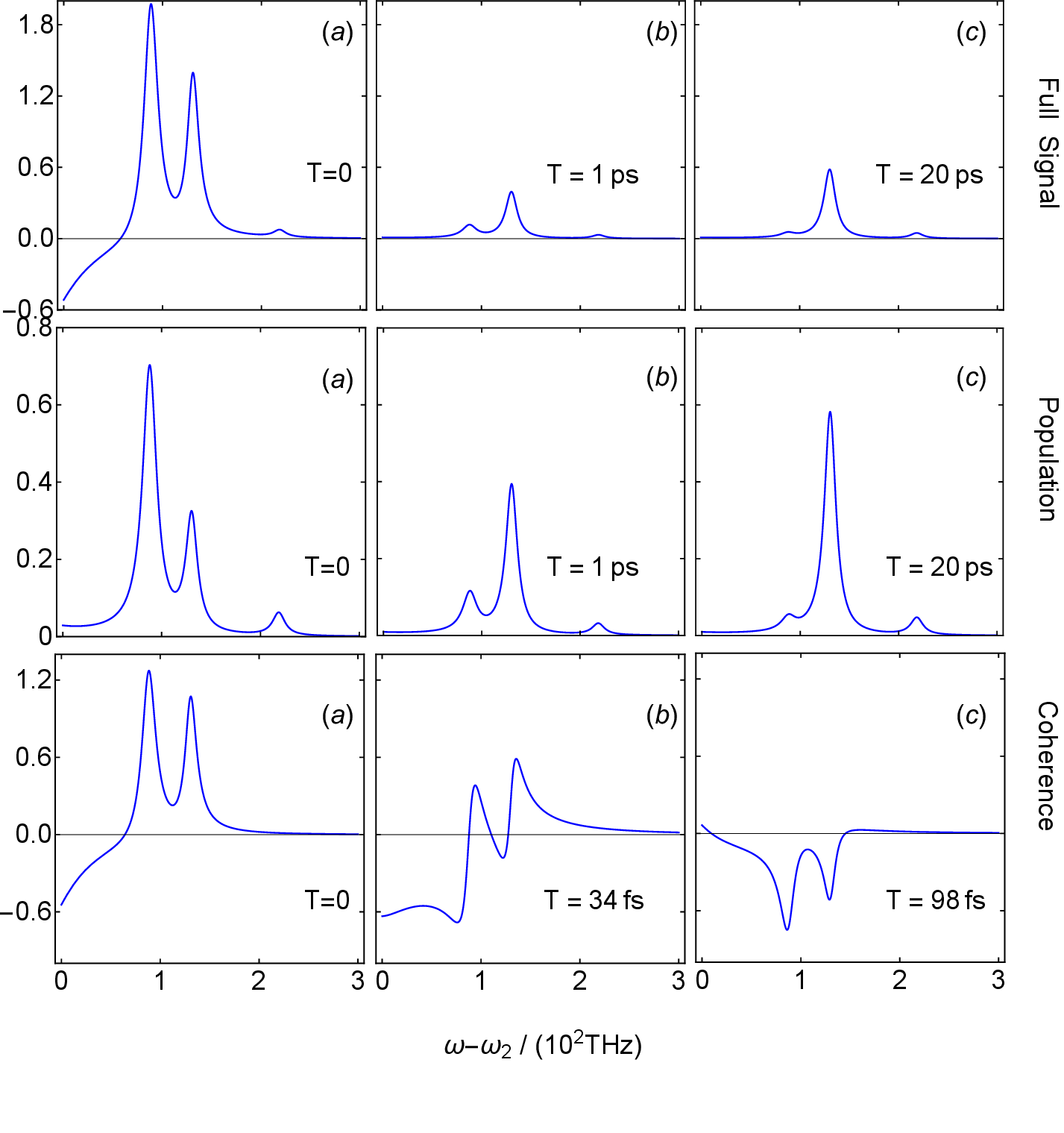}
\caption{The stimulated Raman signals of $N=10$ molecules with a detuning between the molecule and cavity $\omega-v=1.25g$ are presented. Figures (a), (b), and (c) display the stimulated Raman signals at different time delays. Vertically, the total signals, populations, and coherence are displayed sequentially. The relative intensity of peaks observed during real-time delay can be used to resolve the populations of the molecular states.}
\label{1dtot+}
\end{figure}

\begin{figure}[h]
\includegraphics[width=0.35\textwidth]{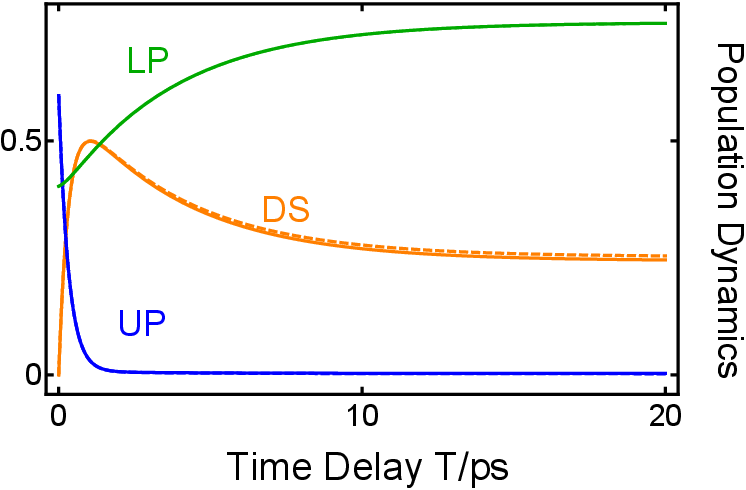}
\caption{Comparison between the populations resolved by the Raman signal (dashed line) and the populations obtained from the master equation (solid line) with a detuning of $1.25g$. The transition dipole is $\mu_{\rm{g,up}}=2.443,\mu_{\rm{g,lp}}=2.008$. The green, orange, and blue colors correspond to the dynamics of the upper polariton, dark state, and lower polariton, respectively.}
\label{1dreslove+}
\end{figure}

Beginning with the analysis of the population shown in the middle line of FIG. \ref{1dtot+}, we clearly observe three peaks at $\omega_{\rm up,ds}=88$THz, $\omega_{\rm ds,lp}=130$THz, and $\omega_{\rm up,lp}=213$THz. As in the previous analysis, the peak at $\omega_{\rm up,ds}$ is expected to represent the combined dynamics of UP and dark states. This peak decreases with time delay, contrasting with the behavior of the LP state. The peak at $\omega_{\rm{ds,lp}}$, however, continually increases, which aligns with the growth of the LP state. This peak corresponds to the combined dynamics of the LP and dark states. Similarly, the peak for $\omega_{\rm up,lp}$ reflects the dynamics of UP and LP states. Its behavior is consistent with the dynamics of the superposition of UP and LP, which first decreases and then begins to increase.
The population dynamics of polaritons coupled to the dark states can be read out from the Raman signal. For a longer timescale, longer than the coherence lifetime, the population dynamics of cavity polaritons dominates. After some manipulations, we are able to find a group of algebraic equations for populations and signal  under impulsive approximation, up to
small errors, i.e.,
\begin{widetext}
\bea
&&S_{\rm 1D}^{(p)}(\omega_{\rm up}-\omega_{\rm ds},T)\simeq\tilde{\alpha}^{(1)}_{\rm 1D,upds}(\rho_{\rm dsds}(T)
+\rho_{\rm upup}(T))+\tilde{\alpha}^{(1)}_{\rm 1D,dslp}(\rho_{\rm dsds}(T)
+\rho_{\rm lplp}(T))
+\tilde{\alpha}^{(1)}_{\rm 1D,uplp}(\rho_{\rm lplp}(T)+\rho_{\rm upup}(T))\,,
\cr && \cr &&
S_{\rm 1D}^{(p)}(\omega_{\rm ds}-\omega_{\rm lp},T)\simeq\tilde{\alpha}^{(2)}_{\rm 1D,upds}(\rho_{\rm dsds}(T)
+\rho_{\rm upup}(T))+\tilde{\alpha}^{(2)}_{\rm 1D,dslp}(\rho_{\rm dsds}(T)
+\rho_{\rm lplp}(T))
+\tilde{\alpha}^{(2)}_{\rm 1D,uplp}(\rho_{\rm lplp}(T)+\rho_{\rm upup}(T))\,,
\cr && \cr &&
S_{\rm 1D}^{(p)}(\omega_{\rm up}-\omega_{\rm lp},T)\simeq\tilde{\alpha}^{(3)}_{\rm 1D,upds}(\rho_{\rm dsds}(T)
+\rho_{\rm upup}(T))+\tilde{\alpha}^{(3)}_{\rm 1D,dslp}(\rho_{\rm dsds}(T)
+\rho_{lplp}(T))+\tilde{\alpha}^{(3)}_{\rm 1D,uplp}(\rho_{\rm lplp}(T)+\rho_{\rm upup}(T))\,
\cr &&\label{eq: resolve1}
\eea
\end{widetext}
where the coefficients are defined by
\bea
\tilde{\alpha}^{(j)}_{{\rm1D},ab}=\frac{1}{\pi} {\rm Im}(-i)^{3}\sum_{n}^{N-1}|\alpha_{a,b}|^2
f(\omega^{(j)},\xi_{ab})
\,\nonumber\\
\eea
Here we define the function $f(v,\xi)$ to simplify the final expressions, which is a function related only to the pulse parameter and the energy gap and derived from integral of \eqref{1dp}:
\begin{align}
&f(v,\xi)=\ft{1}{{-\fft{1}{\sigma_{2}}}+{i \delta }
{-\fft{1}{\sigma_{3}}} 
+{i{{\xi}}}} \Big(\ft{1}{i (v-{\xi})-\fft{1}{\sigma_{2}}}-\ft{1}{{{-\fft{2}{\sigma_{2}}}+{i(v+ \delta) }
{-\fft{1}{\sigma_{3}}} 
}}\Big)\,
\end{align}
where we assume the broadband Raman pulse $\varepsilon_3$ has duration much shorter than the polariton relaxation and dephasing process.
In these formulas, $j=1,2,3$ indicates that $\omega^{(1,2,3)}$ corresponds to $\omega_{\rm up,ds},\omega_{\rm ds,lp},\omega_{\rm up,lp}$ respectively, and $\delta=\omega_2-\omega_3$. Using \eqref{eq: resolve1}, we can solve for the density matrices $\rho_{\rm{ds,ds}}$, $\rho_{\rm{up,up}}$ and $\rho_{\rm{lp,lp}}$ in terms of the Raman spectral lines,  given superposition $\mathcal{N}^{-1}(\mu_{\rm up,g}|\rm UP\rangle+\mu_{lp,g}|\rm LP\rangle)$ an initial state. { We illustrate the dynamics of this picture in FIG. \ref{1dreslove+}, where the initial state is a superposition of LP and UP states with $\mu_{\rm up,g}=2.443, \mu_{\rm lp,g}=2.008$, and the normalization is given by $\mathcal{N}\simeq 3.162$. Due to the existing detuning between molecules and the cavity, the transition dipole between UP and LP is no longer equal, which explains why UP and LP states have different initial weights in FIG. \ref{1dreslove+}.} As displayed in FIG. \ref{1dreslove+}, the bright and dark polariton populations resolved by Eq.(\ref{eq: resolve1}) with the Raman signal have a perfect match to the ones obtained from the Redfield equation Eq.(\ref{master}).  Interestingly, after relaxation ends around $T=20$ps, dark-state polaritons occupy a significant proportion. This insightful finding suggests that dark states can be more effectively explored after relaxation.

Although we use Raman spectroscopy to accurately resolve the dynamics of the upper and lower polaritons, as well as the dark states, the main restriction is that we can only initially pump the superposition of the upper and lower polaritons. We cannot initially pump to any of them individually. Fortunately, we can adjust the time delay between $\varepsilon_1$ to precisely pump either the upper polariton or the lower polariton.

\section{Two-dimensional UV femtosecond stimulated Raman spectroscopy}
\label{sec: result}

To overcome the spectral bottleneck by the 1D Raman spectra above, we will use two pulses for a selective excitation of molecular polaritons. Here a pair of short pulses with an additional delay $T_0$ are pumping the system, creating resonant excitations. The Raman emission is collected after a delay of $T$ relative to the 2nd pump pulse, as depicted in FIG. \ref{mp}(b). Using the Dyson series up to the 2nd order against the couplings with the resonant pump pulses, we find the polariton density matrix
\begin{equation}
    \begin{split}
        \rho(t) & = N \iint_{-\infty}^{t} d\tau_2 d\tau_1 \theta(\tau_2-\tau_1)  \mu(\tau_2)\rho_0 \mu(\tau_1) \\[0.2cm]
        & \qquad\quad \times  \varepsilon_1(\tau_2 - T_1^\prime) \varepsilon_1^*(\tau_1 - T_1) + \text{h.c.} \\[0.2cm]
        & \approx N \int_{-\infty}^{\infty} dt' \int_{-t'}^{\infty} dt'' \hat{G}(t-T_1^\prime) [\mu(t')\rho_0\mu(-t'')] \\[0.2cm]
        & \qquad\qquad\qquad \times \varepsilon_1(t') \varepsilon_1^*(T_0-t'') + \text{h.c.}
    \end{split}
\label{rhop}
\end{equation}
given $t\gg$ pulse duration. $\varepsilon_1(t - T_1),\ \varepsilon_1(t - T_1^\prime)$ denote the two resonant pump fields and $T_0\equiv T_1^\prime - T_1$. $\hat{G}(t)$ is the Green's propagator defined in Eq.(\ref{mastersolution}) for the polariton systems. Eq.(\ref{rhop}) enables the definition of the {\it doorway operators} for the resonant excitation process \cite{pra4164851990}
\begin{equation}
  \begin{split}
    & D(T_0) = N \int_{-\infty}^{\infty} dt' \int_{-t'}^{\infty}dt'' \varepsilon_1(t') \varepsilon_1^*(T_0-t'') \\[0.2cm]
    & \qquad\qquad\qquad\qquad \times \mu(t')\rho_0 \mu(-t'') + \text{h.c.}, \\[0.2cm]
    & D(\omega_0) = \int_{-\infty}^{\infty} d T_0 e^{i\omega_0 T_0} D(T_0)
  \end{split}
\end{equation}
so that
\begin{equation}
    \rho(t) = \hat{G}(t-T_1^\prime) D(T_0).
\label{rhoD}
\end{equation}
Inserting Eq.(\ref{rhoD}) into Eq.(\ref{1dp}) and performing the Fourier transform over $T_0$, the calculations proceed as usual. We thus obtain the 2DUV-FSRS signal in the {\it doorway-Raman-window} formalism, i.e., 
\begin{equation}
    \begin{split}
        S_{\text{2D}} & (\omega-\omega_2,T,\omega_0) \\[0.2cm]
        & \propto \Im \text{Tr}\Big\{ W(\omega-\omega_2) \big[V(\sigma), \hat{G}(T) D(\omega_0) \big]\Big\}
    \end{split}
\label{2dp}
\end{equation}
with $T \gg$ pulse durations meaning that the Raman pump-probe fields are temporally separated from the resonant pump fields; $W(\omega-\omega_2)$ and $V(\sigma)$ are the Raman window operators given by Eqs.(\ref{RW}) and (\ref{RV}), respectively.
The salient feature of the two-dimensional Raman signal can be easily seen from \eqref{2dp}: it allows one to selectively access the polariton states by the pump. We can then expect a real-time monitoring of the pathways of the polariton dynamics, which are a hard task for the 1D Raman signal.

\subsection{Without detuning $\omega-v=0$}

\begin{figure}[h]
\includegraphics[width=0.45\textwidth]{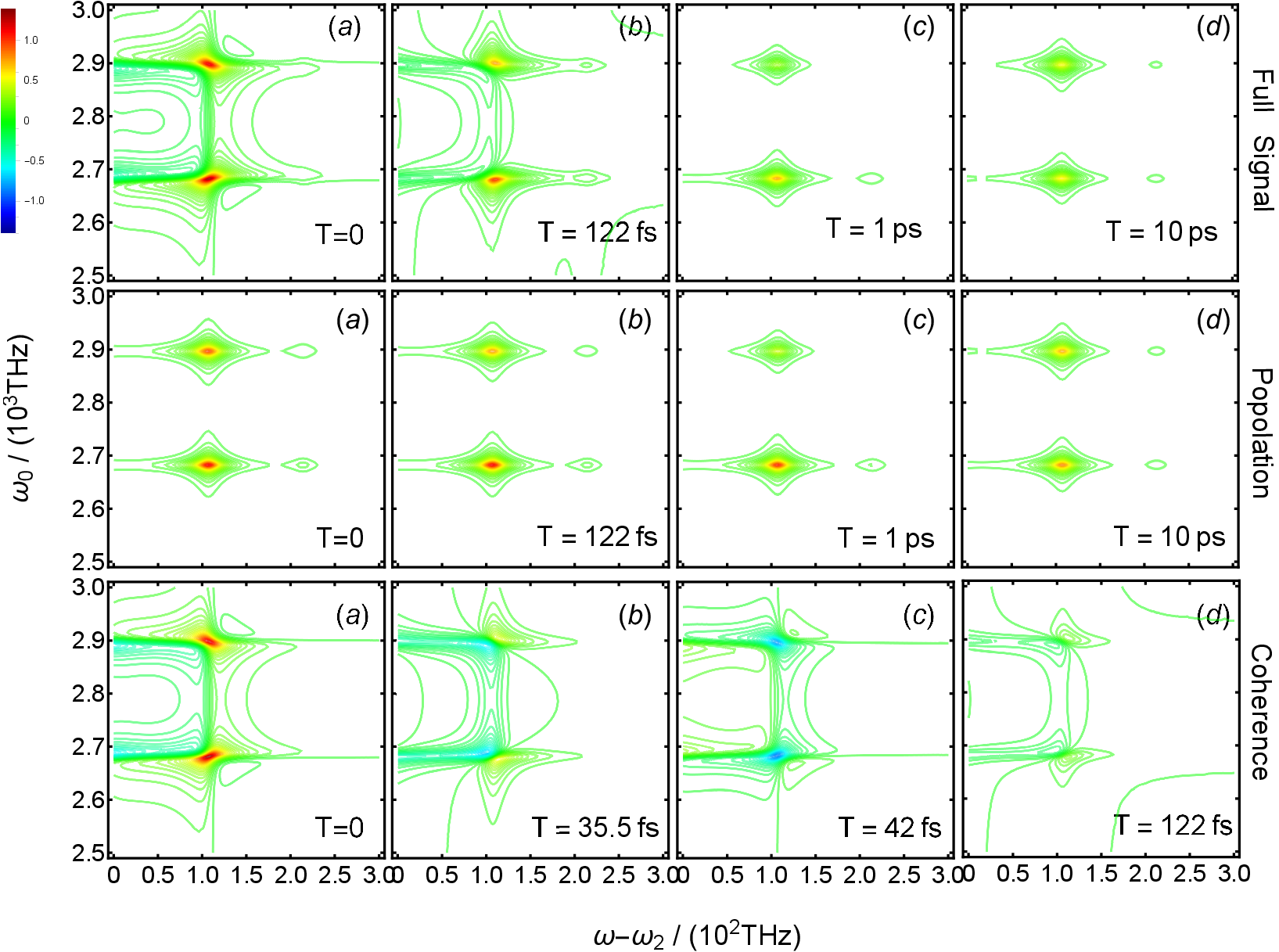}
\caption{Two-dimensional Raman signal without the detuning.}
\label{2dtot0}
\end{figure}

Let's briefly analyze  the behavior of the 2D Raman signal shown in FIG. \ref{2dtot0}, with $\delta = 0$. Different values of $\omega_0$ allow us to select different initial pumping states, as indicated by the vertical axis of FIG. \ref{2dtot0}. From the vertical axis, it is clear that the peaks are localized around either the upper polariton ($\omega_0=2890$THz) or the lower polariton ($\omega_0=2680$THz), referring to the initial pumping to the upper polariton and lower polariton, respectively. In other words, we expect that analyzing the peaks that vertically localize around the upper polariton will reveal the real-time populations for the initial upper pump, and the same applies to the lower polariton case.

To analyze the case with the initial pumping to the upper polariton, we should focus on the upper peaks in a given plot of FIG. \ref{2dtot0}. We observe that the signal is dominated by two peaks corresponding to $\omega-\omega_2=\omega_{\rm{up,ds}}=\omega_{\rm{ds,lp}}=103$THz and $\omega-\omega_2=\omega_{\rm{up,ds}}=206$THz. Because $\omega-v=0$, the energy gap $\omega_{\rm{up,ds}}$ equals $\omega_{\rm{ds,lp}}$. In this case, the first peak is the merged one from $\omega_{\rm{up,ds}}$ and $\omega_{\rm{ds,lp}}$ due to zero detuning. Consequently, as in the one-dimensional case, the peak at $\omega-\omega_2=\omega_{\rm{up,ds}}=\omega_{\rm{ds,lp}}$ reflects a mixed population from the upper polariton, dark state, and lower polariton with different ratios, making it difficult to obtain valid information about each of them. However, the peak at $\omega-\omega_2=\omega_{\rm{up,ds}}$ only encodes the information of the upper and lower polariton. Its intensity first decreases and then grows, which is consistent with the dynamics of $\rho_{\rm{up,up}}+\rho_{\rm{lp,lp}}$ shown in FIG. \ref{resolve10}(a). Similarly, when the initial pump is pulsed to the lower polariton (the $\omega_0=2680$THz line), the intensity of the peak at $\omega-\omega_2=\omega_{\rm{up,ds}}$ continues to decrease, which aligns with the dynamics of $\rho_{\rm{up,up}}+\rho_{\rm{lp,lp}}$, as shown in FIG. \ref{resolve10}(b).

Regarding the coherence part of FIG. \ref{2dtot0}, we see the peaks are concentrated vertically on the upper polariton ($\omega_0=2890$THz) and lower polariton ($\omega_0=2680$THz), implying that the coherence oscillates only between the UP and LP states. A simple explanation is that we cannot initially pump to the dark state. We can clearly observe the oscillation and decay in the coherence section of FIG. \ref{2dtot0}. The combination of population and coherence is shown on the top line in FIG. \ref{2dtot0}. We observe that there are no coherence contributions anymore at $\rm{T}=1\rm{ps}$.

Until now, we have observed the dynamics of $\rho_{\rm{up,up}}+\rho_{\rm{lp,lp}}$. However, obtaining their individual dynamics is challenging due to the overlapping of the first two peaks $\omega-\omega_2=\omega_{\rm{up,ds}}$ and $\omega-\omega_2=\omega_{\rm{ds,lp}}$. In the following sections, we will turn on the detuning, which will separate these two peaks.

\subsection{With detuning $\omega-v=1.25g$}

\begin{figure}[h]
\includegraphics[width=0.45\textwidth]{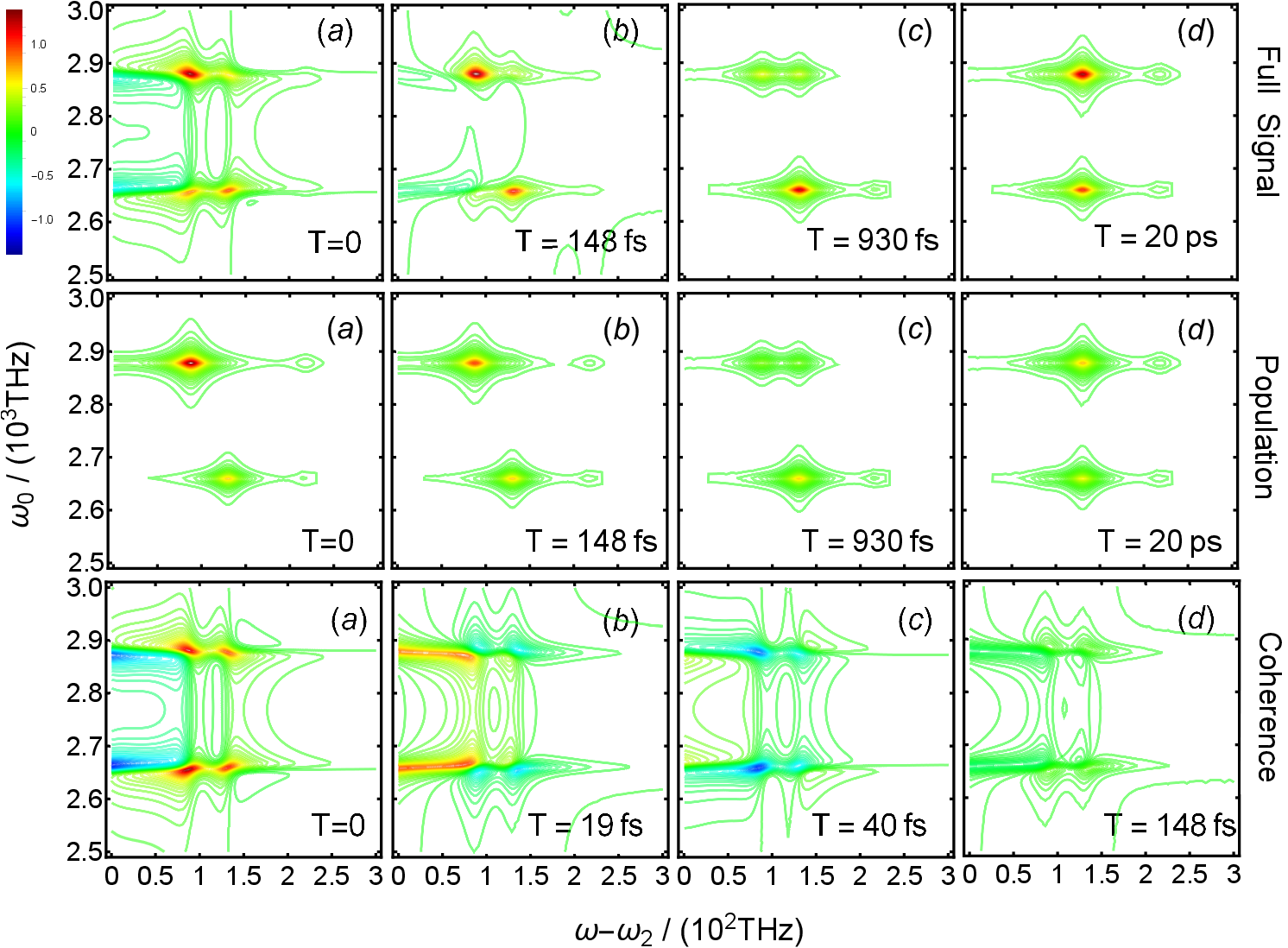}
\caption{Two-dimensional Raman signal obtained with a detuning of $1.25g$.}
\label{2dtot+}
\end{figure}

In a global view, we observe two peaks at $\omega_{\rm{up}}$ and $\omega_{\rm{lp}}$ when taking slices along $\omega_0$. This is due to the destructive interference in the dark-state polaritons that results in zero net dipole. Nevertheless, three peaks can be seen along the slices with fixed $\omega_0$. This indicates the active response of the dark-state polaritons.

The 1st row of FIG. \ref{2dtot+} shows the full 2DUV-FSRS given by Eq.(\ref{2dp}), with scanning the delay T. For the slice at $\omega_0=\omega_{\rm{up}}=2890$THz, FIG. \ref{2dtot+}(a) depicts that the signal is dominated by two peaks at $\omega_{\rm{up,ds}}=88$THz and $\omega_{\rm{up,lp}}=218$THz. This means the population of the system at the upper polariton state. When the delay $\rm{T}$ varies, we see the energy transfer from UP to DSPs, evident by the increase of the peak at $\omega-\omega_2=\omega_{\rm{ds,lp}}$ in FIG. \ref{2dtot+}(b) and \ref{2dtot+}(c). After a longer delay, as seen from FIG. \ref{2dtot+}(d), the DSPs are densely populated whereas the population at LP state is less. This is attributed to the large number of the DSPs.

FIG. \ref{2dtot+}(a-d) further shows different timescales associated with different pathways. In particular, the peak at $(\omega-\omega_2=\omega_{\rm{ds,lp}}, \omega_0=\omega_{\rm{up}})$ elaborates a fast increase within 930fs, dramatically different from the one at ($\omega-\omega_2=\omega_{\rm{up,lp}}, \omega_0=\omega_{\rm{up}}$) that shows a considerable change within 20ps. We thus observe a faster energy transfer from UP to DSPs than that from UP to LP. This can be understood neatly by the large density of the DSPs, much higher than the bright polariton states that have been revealed in resonant spectroscopic experiments. The transition rate is {{$\frac{2 \pi}{\hbar}\langle \rm DS|V_{\rm int}| \rm UP\rangle|^2 \rho(E_{ds})$}} from the Fermi's Golden rule, which is enhanced by the mode density of the states.

When slicing FIG. \ref{2dtot+}(a-d) at $\omega_0=\omega_{\rm{lp}}$, we see the dynamics when initially pumping to the LP state. The results present dramatical difference from the ones with $\omega_0=\omega_{\rm{up}}$, resulting in a bit more subtle analysis due to the resolution issue for the figures. Nonetheless, we are still able to find the peak around $\omega_{\rm{ds,lp}}$ barely varying with the delay, while the spectral line got broadened slightly towards the the peak at $\omega_{\rm{up,ds}}$. The peak at $\omega-\omega_2=\omega_{\rm{up,lp}}$ shows a low intensity, which  indicates a weak population at the UP state.

Notably from the comparison between the top and middle rows of FIG. \ref{2dtot+}, the DSPs can be greatly populated when the system is pumped to the UP rather than the LP state. This indicates the energy harvesting by the DSPs, which is important for understanding the kinetic and thermodynamic properties of molecules in cavities. As an optical signal, FIG. \ref{2dtot+} clearly demonstrates a real-time monitoring of the DSPs coupled to the other states of molecules, for an illustration of the crucial role of the DSPs.

Within shorter timescale, the 2DUV-FSRS reveals the coherence effect in the cavity-polariton systems, evident by the fast oscillations. By a careful check with the oscillating frequency, as supported by FIG. \ref{2dtot+}(3rd row), the quantum coherence between UP and LP states is captured. This is due to the broadband nature of the pump fields that create a coherent superposition $a|\rm{UP}\rangle + b|\rm{LP}\rangle$. Such a polaritonic coherence is resolved by the Raman signal during a short timescale, whereas the Raman signal is dominated by the polariton populations during longer timescales.

\begin{figure}[h]
\includegraphics[width=0.45\textwidth]{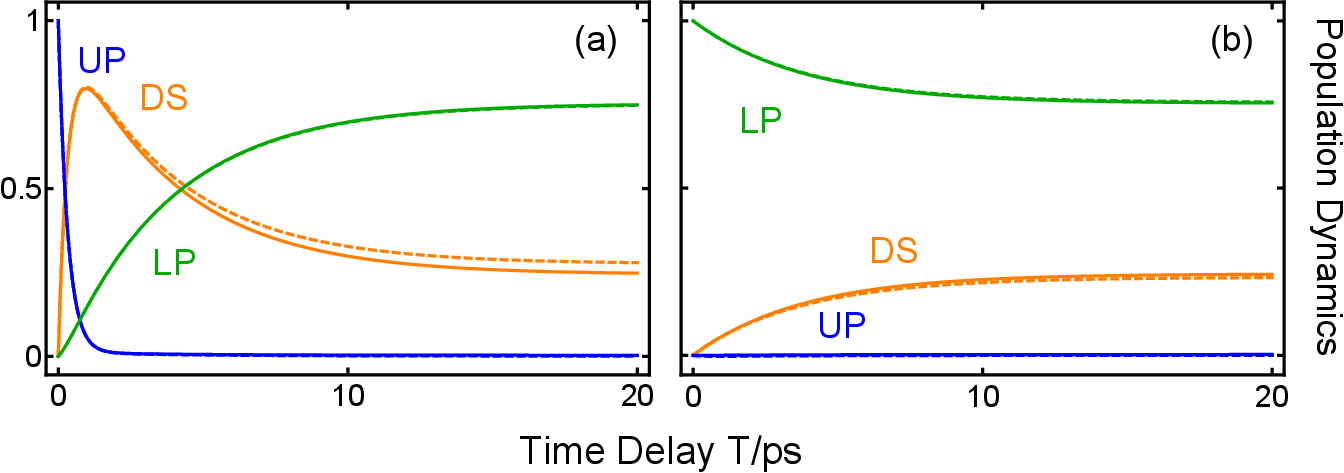}
\caption{Comparison between the population resolved by the signal (dashed line) and the population obtained from the master equation (solid line), for $\omega-v=1.25g$.
 }
\label{resolve10}
\end{figure}

The polariton dynamics may be monitored in a more advanced way, through a sophisticated method, i.e., the intrinsic connection of peak intensities to the populations that dominate over longer timescales. A group of algebraic equations can thus be found
\begin{widetext}
\bea
&&S_{\rm 2D}^{(p)}(\omega_{\rm up}-\omega_{\rm ds},\omega_{\rm up},T)\simeq\tilde{\alpha}^{(1)}_{\rm up,ds}(\rho_{\rm dsds}(T)
+\rho_{\rm upup}(T))+\tilde{\alpha}^{(1)}_{\rm ds,lp}(\rho_{\rm dsds}(T)
+\rho_{\rm lplp}(T))
+\tilde{\alpha}^{(1)}_{\rm up,lp}(\rho_{\rm lplp}(T)+\rho_{\rm upup}(T))\,,
\cr && \cr &&
S_{\rm 2D}^{(p)}(\omega_{\rm ds}-\omega_{\rm lp},\omega_{\rm up},T)\simeq\tilde{\alpha}^{(2)}_{\rm up,ds}(\rho_{\rm dsds}(T)
+\rho_{\rm upup}(T))+\tilde{\alpha}^{(2)}_{\rm ds,lp}(\rho_{\rm dsds}(T)
+\rho_{\rm lplp}(T))
+\tilde{\alpha}^{(2)}_{\rm up,lp}(\rho_{\rm lplp}(T)+\rho_{\rm upup}(T))\,,
\cr && \cr &&
S_{\rm 2D}^{(p)}(\omega_{\rm up}-\omega_{\rm lp},\omega_{\rm up},T)\simeq\tilde{\alpha}^{(3)}_{\rm up,ds}(\rho_{\rm dsds}(T)
+\rho_{\rm upup}(T))+\tilde{\alpha}^{(3)}_{\rm ds,lp}(\rho_{\rm dsds}(T)
+\rho_{\rm lplp}(T))+\tilde{\alpha}^{(3)}_{\rm up,lp}(\rho_{\rm lplp}(T)+\rho_{\rm upup}(T))\,,\label{reseq}
\cr && \label{eq: resolve 2D}
\eea
\end{widetext}
where the coefficients are defined by
\bea
\tilde{\alpha}^{(j)}_{a,b}&=&\frac{1}{\pi} {\rm Im}(-i)^{3}\sum_{n}^{N-1}|\alpha_{a,b}|^2|\mu_{\rm g,up}|^2\times
\cr&&
(\fft{1}{i \gamma_{g{up}}}+\fft{1}{i\gamma_{{up}g}})f(\omega^{(j)},\xi_{ab})
\,
\eea
We can solve the density matrices from \eqref{eq: resolve 2D} in both cases, i.e., when pumping to the upper and lower polariton. This resolved population turns out to match perfectly with the real dynamics governed by the master equation (see FIG. \ref{resolve10}). 

\subsection{With detuning $\omega-v=-1.25g$}

\begin{figure}[h]
\includegraphics[width=0.45\textwidth]{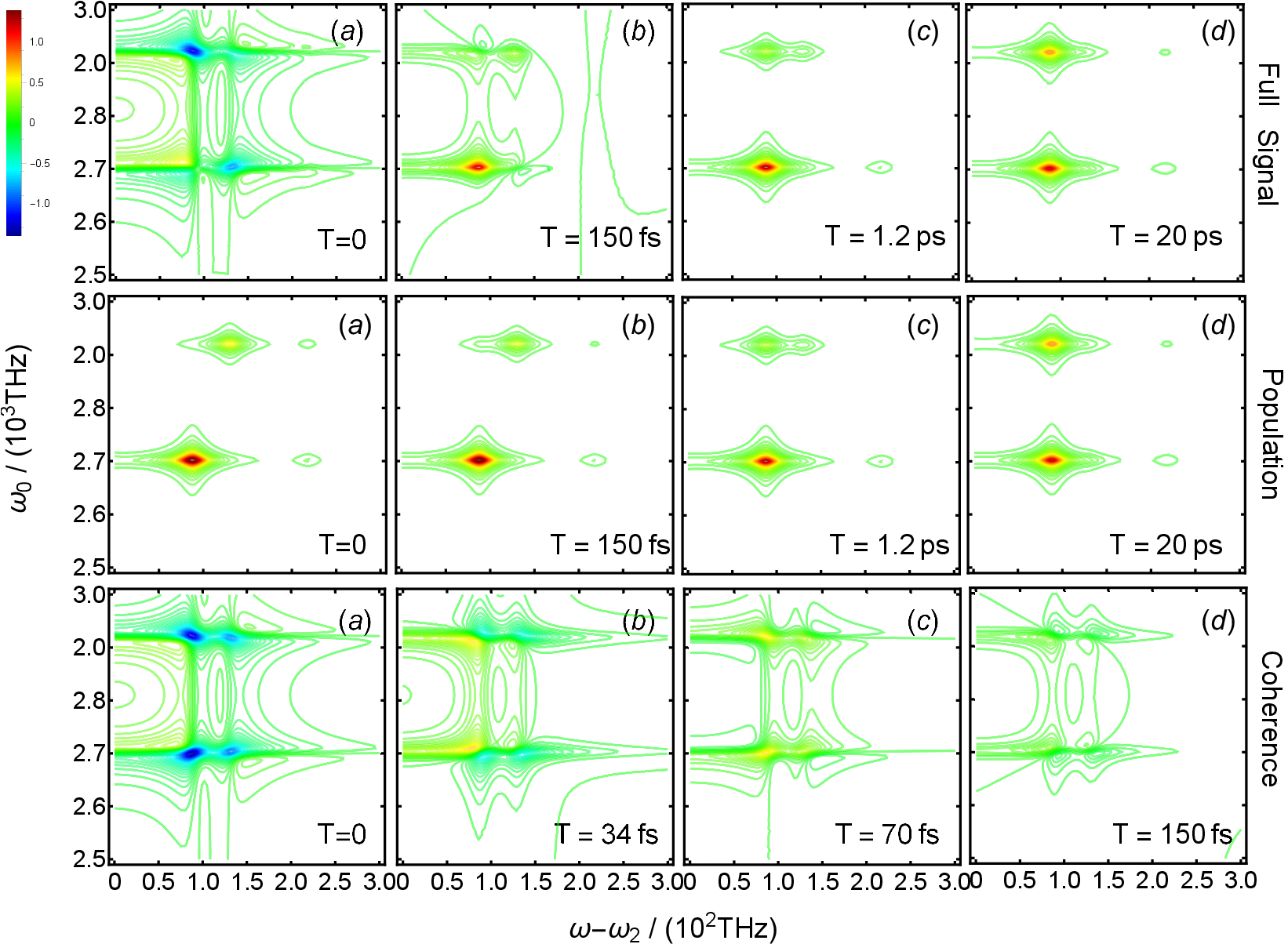}
\caption{Two-dimensional Raman signal with the detuning of $\omega-v=-1.25g$.}
\label{2dtot-}
\end{figure}

We can also consider the negative detuning, such as $\omega-v=-1.25g$. In this case, $\omega_{\rm{up,lp}}$ is the same as in the previous case (i.e., $\omega-v=1.25g$); however, the energy gaps of $\omega_{\rm{up,ds}}$ and $\omega_{\rm{ds,lp}}$ have been exchanged. Similarly, we show the full signal, population, and coherence all in FIG. \ref{2dtot-}.

\begin{figure}[h]
\includegraphics[width=0.45\textwidth]{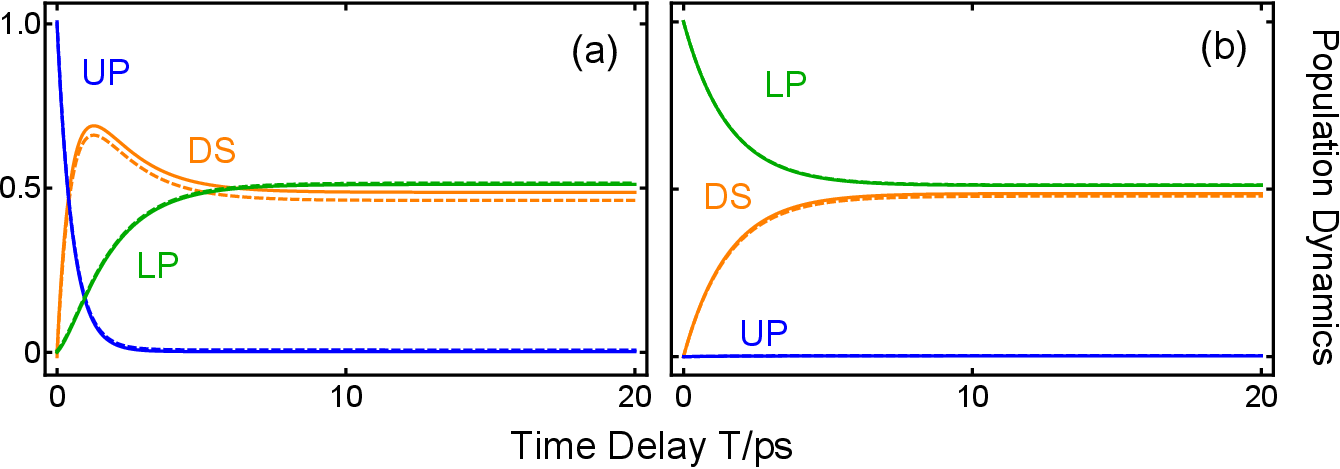}
\caption{Comparison between the population resolved by signal (dashed line) and the population obtained from master equation (solid line) for $\omega-v=-1.25g$.}
\label{reslove10-}
\end{figure}

It is immediately evident from FIG. \ref{2dtot-} that the trends for the peaks do not deviate significantly from the positive detuning case, indicating similar behavior for the populations, which we do not need to discuss further. However, the specific details of the dynamics indeed change. We demonstrate these changes in the resolved dynamics and the real dynamics in FIG. \ref{reslove10-}, which, of course, fit well. The most significant discrepancy with the positive detuning is a substantial increase in the percentage of the dark state in the populations. This occurs because dark states with negative detuning have lower energy comparing to the positive detuning case, making the dark states more easily ``accessible''. Therefore, to prevent the dark state from becoming scarce, we could consider setting the detuning to a negative value.

\section{2DUV-FSRS with charge transfer states}
\label{sec: ct}
There are molecules reactive for electron transfer, accessing the charge transfer states (CTs). The CTs normally have energies typically lower than the first excited states of the molecules \cite{chargestate}. Our strategy, basically, is to induce the stimulated Raman transition that ends up at the CTs. As the CTs are hard to be excited directly by the pumping pulse, the Raman signal for the molecular polaritons can be cleaner than before. This is revealed by the fact that the components (a) and (c) in the loop diagrams (FIG. \ref{loop}) survive only. 
Assuming the near-edge states can coupled radiatively to both the molecular excitons and the CTs, the Raman polarizability for the molecular polariton reads the form
Eq.\eqref{34}
following Eq.(\ref{35})
\begin{equation}
    \alpha_{{e_3}{ct}}=\sum_{i}^{N} \ft{P_{i}\mu_
    {ri,ct}U_{i {e_3}}^\dagger }{\hbar}\left(\ft{1}{\omega_{i}-\omega_{{ct}}}+\ft{1}{\omega_{i}-\omega_{{e_3}}}\right)
\label{34}
\end{equation}
where $\mu_{ri,ct}$ denotes the transition dipole between the near edge state and the charge transfer state. For this system, we find the 2DUV-FSRS signal with CTs 
\begin{align}
    &S_{ct}(\omega_0,\omega-\omega_2,T)=\fft{1}{\pi}{\rm Im}(-i)^{3}\varepsilon_3^*(\omega)\int^\infty_0 dt\int^t_0 d\tau\nonumber \\
    & \sum_{e_1e_2e_3e_4}^{N+1}\mu_{ge_1}\mu_{ge_2}\alpha_{e_3ct}\alpha_{cte_4}(\fft{1}{-\omega_0-\xi_{ge_1}}+\fft{1}{\omega_0-\xi_{e_1g}})\nonumber \\
    &\varepsilon_2(t-T)
    e^{i\omega(t-T)}\varepsilon_2^*(\tau-T)\varepsilon_3(\tau-T)G_{e_4e_3,e_2e_1}(\tau-T_2)\nonumber \\
    &e^{-i\omega_{e_4ct}(t-\tau)}\label{ctp}
\end{align}

\begin{figure}[h]
\includegraphics[width=0.45\textwidth]{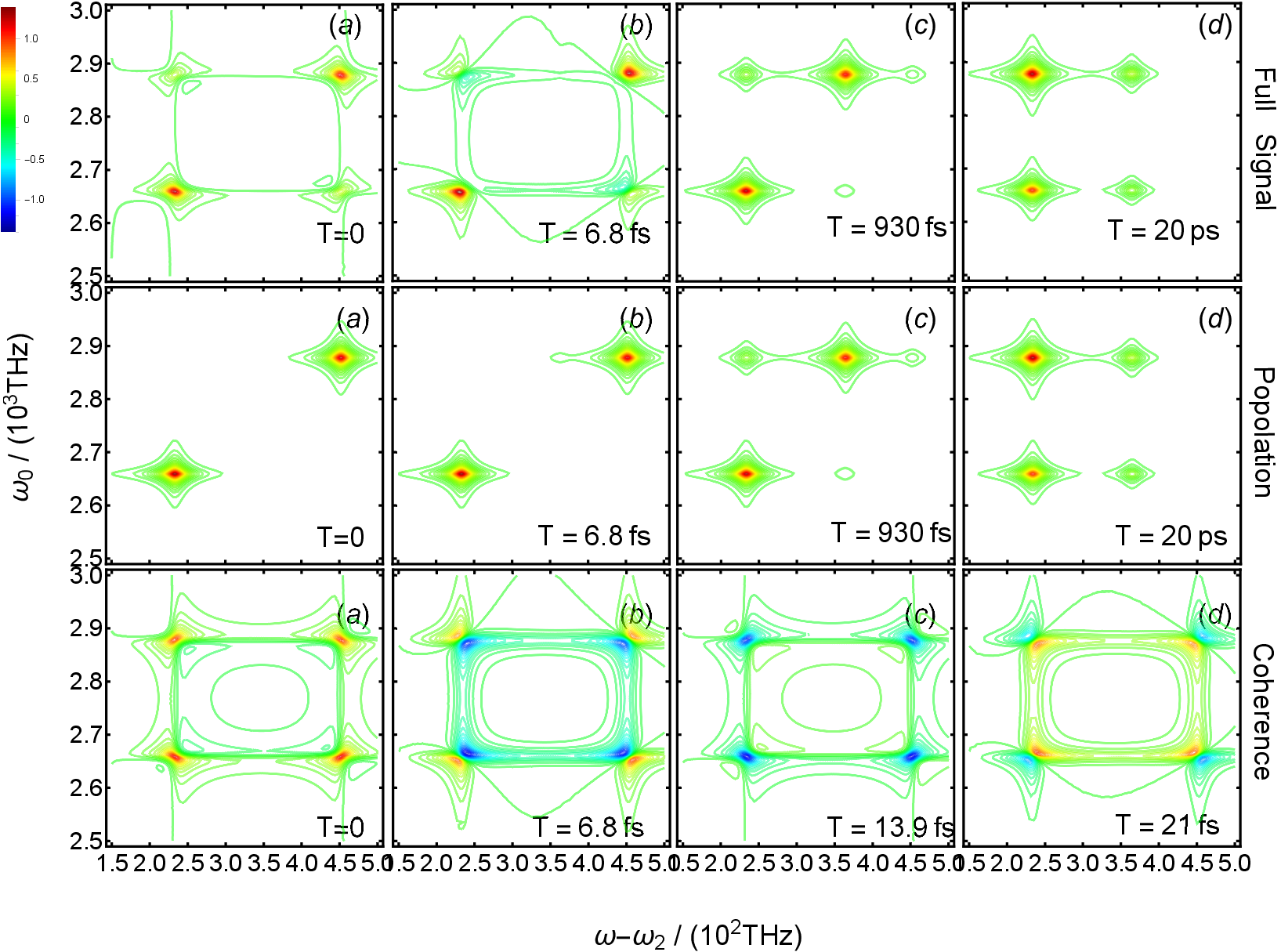}
\caption{This is the Raman signal with detuning $\omega-v=1.25g$, and the energy of the charge transfer state is $1.6$eV.}
\label{2dloctot}
\end{figure}

For visualizing our signal, we consider 10 molecules and assume that the energy of the charge transfer state is $1.6$eV, as shown in FIG. \ref{2dloctot}. Because the parametric process is not included in this case, a peak located at $\omega_{e_3e_5}\geq 0$ perfectly reflects the density matrix $\rho_{e_3,e_3}$. This fact should be remembered when analyzing FIG. \ref{2dloctot} and comparing it with the real-time dynamics in FIG. \ref{resolve10}. First, we consider the pulse initially pumping to the upper polariton, which is shown in the upper part of the signal plots in FIG. \ref{2dloctot}. The behaviour of the peak at $\omega-\omega_2=\omega_{\rm{lp,ct}}=230$THz shows the dynamics of the upper polariton, which continues to grow as indicated by the increase in the intensity of the relevant peak. On the other hand, the peak at $\omega-\omega_2=\omega_{\rm{ds,ct}}=360$THz encodes the dynamics of the dark state: the density matrix increases over time and then starts to decay at some point. Lastly, the peak at $\omega-\omega_2=\omega_{\rm{up,ct}}=450$THz encodes the dynamics of the upper polariton, which keeps decreasing as reflected by the decreasing peak intensity.

Regarding the lower part of the population in FIG. \ref{2dloctot}, where the system is initially pumped to the lower polariton, the peak at $\omega-\omega_2=\omega_{\rm{up,ct}}$ is absent for all time delays because the energy of the upper polariton is too high to undergo relaxation. The decrease in intensity for the peak at $\omega-\omega_2=\omega_{\rm{lp,ct}}$ and the opposite trend in intensity for the peak at $\omega-\omega_2=\omega_{\rm{ds,ct}}$ correspond to the decay of the UP state and the growth of the dark state, respectively.

For the coherence part of FIG. \ref{2dloctot}, we observe that peaks primarily dominate around $\omega-\omega_2=\omega_{\rm{lp,ct}}$ and $\omega-\omega_2=\omega_{\rm{up,ct}}$. This is due to the selection rule that forbids the initial pumping to the dark states. As usual, the peaks display an oscillatory decay. Moreover, we can clearly see the oscillation at the peak, transitioning from ``all red", through ``half red, half blue", to ``all blue". As the color fades, it also shows that the oscillation is in decay. For the total signal, it is worth noting that the ``red dot" (positive part) in FIG. \ref{2dloctot}(b) always represents the contribution of the population because the positive and negative values of coherence are the same. We can see that FIG. \ref{2dloctot}(c) and (d) are the same as the pure population, implying that the coherence has dissipated.

\section{Conclusion and remarks}
\label{sec:conclu}
We studied the coherent Raman response of the molecular polaritons, in which the dark states are visualized. The results led to the 2DUV-FSRS for cavity-polariton systems, and we therefore developed a microscopic theory for the Raman signal. Rich information about the dark-state polaritons and their coupling to the bright polaritons can be readily visualized in the 2DUV-FSRS. Our work provides an off-resonant spectroscopic scheme for a real-time monitoring of the dark-state-polariton dynamics, not accessible by conventional spectroscopic technique including the absorption and fluorescence. The multidimensional projections of the Raman signal as elaborated enable a multiscale illustration for the polariton dynamics in a crosstalk with the DPSs, underlying a time- and frequency-resolved nature.

Our work would be insightful for the study of polariton-afforded reactivity of photo-active molecules, and the cavity-coupled heterostructures including the 2D semiconductors. The Raman spectra in present work can enable a clean fingerprint for the fast nonadiabatic electron dynamics, where complex energy potentials involving anharmonicity needs to be taken into account. These would be a remarkable generalization of our present work, and will be presented elsewhere.



\begin{references}
\bibitem{polariton}Fregoni J, Garcia-Vidal F J, Feist J. Theoretical challenges in polaritonic chemistry[J], ACS photonics, (2022).
\bibitem{76-18} Kena-Cohen, S.; Forrest, S. R. Room-temperature polariton lasing in an organic single-crystal microcavity. Nat. Photonics 2010, 4,
371-375.
\bibitem{76-19} Cookson, T.; Georgiou, K.; Zasedatelev, A.; Grant, R. T.;
Virgili, T.; Cavazzini, M.; Galeotti, F.; Clark, C.; Berloff, N. G.;
Lidzey, D. G.; Lagoudakis, P. G. A Yellow Polariton Condensate in a
Dye Filled Microcavity. Adv. Opt. Mater. 2017, 5, No. 1700203
\bibitem{20-3} Plumhof, J. D.; Stoferle, T.; Mai, L.; Scherf, U.; Mahrt, R. F.
Room-Temperature Bose-Einstein Condensation of Cavity Exciton-
Polaritons in a Polymer. Nat. Mater. 2014, 13 (3), 247-252.
\bibitem{42-13} M. A. Sentef, M. Ruggenthaler, and A. Rubio, Cavity
quantum-electrodynamical polaritonically enhanced electron-phonon coupling and its influence on superconductivity, Sci.
Adv. 4, eaau6969 (2018).
\bibitem{42-14}Thomas, Anoop, et al. Exploring superconductivity under strong coupling with the vacuum electromagnetic field. arXiv preprint arXiv:1911.01459 (2019).

\bibitem{20-4} Hutchison, J. A.; Schwartz, T.; Genet, C.; Devaux, E.; Ebbesen,
T. W. Modifying Chemical Landscapes by Coupling to Vacuum
Fields. Angew. Chem. 2012, 124 (7), 1624-1628.
\bibitem{20-5}Thomas, A.; George, J.; Shalabney, A.; Dryzhakov, M.; Varma, S.
J.; Moran, J.; Chervy, T.; Zhong, X.; Devaux, E.; Genet, C.;
Hutchison, J. A.; Ebbesen, T. W. Ground-State Chemical Reactivity
under Vibrational Coupling to the Vacuum Electromagnetic Field.
Angew. Chem., Int. Ed. 2016, 55 (38), 11462-11466.
\bibitem{20-6} Zhong, X.; Chervy, T.; Wang, S.; George, J.; Thomas, A.;
Hutchison, J. A.; Devaux, E.; Genet, C.; Ebbesen, T. W. Non-Radiative Energy Transfer Mediated by Hybrid Light-Matter States.
Angew. Chem., Int. Ed. 2016, 55 (21), 6202-6206.
\bibitem{20-7} Coles, D. M.; Somaschi, N.; Michetti, P.; Clark, C.; Lagoudakis,
P. G.; Savvidis, P. G.; Lidzey, D. G. Polariton-Mediated Energy
Transfer between Organic Dyes in a Strongly Coupled Optical
Microcavity. Nat. Mater. 2014, 13 (7), 712-719.
\bibitem{42-30}  C. Schafer, M. Ruggenthaler, H. Appel, and A. Rubio, Modification of excitation and charge transfer in cavity quantum-electrodynamical chemistry, Proc. Natl. Acad. Sci. USA 116,
4883 (2019).
\bibitem{42-33} Hagenmüller, David, et al. Cavity-assisted mesoscopic transport of fermions: Coherent and dissipative dynamics. Physical Review B 97.20 (2018): 205303.
\bibitem{42-32}Hagenmüller, David, et al. Cavity-enhanced transport of charge. Physical review letters 119.22 (2017): 223601.
\bibitem{42-31} Orgiu, E., et al. Conductivity in organic semiconductors hybridized with the vacuum field. Nature Materials 14.11 (2015): 1123-1129.
\bibitem{JPCA1235918}Xiang, Bo, et al. State-selective polariton to dark state relaxation dynamics. The Journal of Physical Chemistry A123.28 (2019): 5918-5927.
\bibitem{NC}Grafton, Andrea B., et al. Excited-state vibration-polariton transitions and dynamics in nitroprusside. Nature Communications 12.1 (2021): 214.
\bibitem{MarkusJPCA}
Kowalewski, Markus, et al. "Nucleophilic substitution dynamics: Comparing wave packet calculations with experiment." The Journal of Physical Chemistry A 118.26 (2014): 4661-4669.
\bibitem{PRL128096001}Du, Matthew, and Joel Yuen-Zhou. Catalysis by dark states in vibropolaritonic chemistry. Physical Review Letters 128.9 (2022): 096001. 
\bibitem{PRB106L220306}Zhang, Zhedong, Shixuan Zhao, and Dangyuan Lei. "Quantum statistical theory for an exciton-polariton condensate: Fluctuations and coherence." Physical Review B 106.22 (2022): L220306.
\bibitem{cd1}
Ribeiro, Raphael F., et al. Polariton chemistry: controlling molecular dynamics with optical cavities. Chemical science 9.30 (2018): 6325-6339.
\bibitem{cd2}
Xiang, Bo, and Wei Xiong. "Molecular vibrational polariton: Its dynamics and potentials in novel chemistry and quantum technology." The Journal of Chemical Physics 155.5 (2021).
\bibitem{ZhangJPCL2019}Zhang, Zhedong, et al. Polariton-assisted cooperativity of molecules in microcavities monitored by two-dimensional infrared spectroscop. The journal of physical chemistry letters 10.15 (2019): 4448-4454.
\bibitem{XiongPNAS2018}Xiang, Bo, et al. Two-dimensional infrared spectroscopy of vibrational polaritons. Proceedings of the National Academy of Sciences 115.19 (2018): 4845-4850.
\bibitem{u}Cordero, Sergio, et al. Effect of the atomic dipole-dipole interaction on the phase diagrams of field-matter interactions: Variational procedure. Physical Review A 105.3 (2022): 033712.
\bibitem{75}Herrera, Felipe, and Frank C. Spano. Theory of nanoscale organic cavities: The essential role of vibration-photon dressed states. ACS photonics 5.1 (2018): 65-79.
2408.

\bibitem{chem2009}Abramavicius D, Palmieri B, Voronine D V, et al. Coherent multidimensional optical spectroscopy of excitons in molecular aggregates; quasiparticle versus supermolecule perspectives[J]. Chemical reviews, 2009, 109(6): 2350
\bibitem{gamma}Novoderezhkin V I, Palacios M A, Van Amerongen H, et al. Energy-transfer dynamics in the LHCII complex of higher plants: modified redfield approach[J]. The Journal of Physical Chemistry B, 2004, 108(29): 10363-10375.
\bibitem{2}D. W. McCamant, P. Kukura, and R. A. Mathies, J. Phys. Chem. A 107,
8208 (2003).
\bibitem{3}S.-Y. Lee, D. Zhang, D. W. McCamant, P. Kukura, and R. A. Mathies, J.
Chem. Phys. 121, 3632 (2004).
\bibitem{5}P. Kukura, D. W. McCamant, S. Yoon, D. B. Wandschneider, and R. A.
Mathies, Science 310, 1006 (2005).
\bibitem{6}P. Kukura, D. W. McCamant, and R. A. Mathies, Annu. Rev. Phys. Chem.
58, 461 (2007).
\bibitem{8}H. Kuramochi, S. Takeuchi, and T. Tahara, J. Phys. Chem. Lett. 3, 2025
(2012).


\bibitem{raman pro}Zhang Z, Peng T, Nie X, et al. Entangled Photons Enabled Time-and Frequency-Resolved Coherent Raman Spectroscopy in Condensed Phase Molecules[J]. arXiv preprint arXiv:2106.10988, 2021.
\bibitem{near}Chenghao Zhang, Martin Gruebele, and Peter G. Wolynes. Surface crossing and energy flow in many-dimensional quantum systems.pnas.2221690120,2023.
\bibitem{fsrs}Dorfman, Konstantin E., Benjamin P. Fingerhut, and Shaul Mukamel. "Time-resolved broadband Raman spectroscopies: A unified six-wave-mixing representation." The Journal of chemical physics 139.12 (2013): 124113.
\bibitem{jagg}Hobson, Peter A., et al. "Strong exciton–photon coupling in a low-Q all-metal mirror microcavity." Applied Physics Letters 81.19 (2002): 3519-3521.
\bibitem{pra4164851990}Yan, Yi Jing, and Shaul Mukamel. "Femtosecond pump-probe spectroscopy of polyatomic molecules in condensed phases." Physical Review A 41.11 (1990): 6485.
\bibitem{chargestate}Fassioli, Francesca, et al. "Femtosecond photophysics of molecular polaritons." The Journal of Physical Chemistry Letters 12.46 (2021): 11444-11459.


\end{references}
\end{document}